\documentclass[1p]{elsarticle}

\usepackage{hyperref}
\usepackage{subcaption}
\usepackage{longtable}
\journal{Chemical Physics Letters}

\usepackage{graphicx}
\usepackage{longtable}
\usepackage{natbib}

\begin{document}

\begin{frontmatter}

\title{A global optimisation study of the low-lying isomers of the alumina octomer (Al$_2$O$_3$)$_8$}

%\author{David Gobrecht\fnref{myfootnote}}
%\address{Celestijnenlaan 200D, 3001 Leuven}

\author[mymainaddress]{David Gobrecht\tnoteref{mytitlenote}}
\tnotetext[mytitlenote]{corresponding author}
\ead{david.gobrecht@kuleuven.be}
%\ead{s.bromley@ub.edu}

\author[mymainaddress]{Leen Decin}

\author[myquaternaryaddress,myquinternaryaddress]{Sergio Cristallo}

\author[mysecondaryaddress,myternaryaddress]{Stefan T. Bromley}

\address[mymainaddress]{Instituut voor Sterrenkunde, KU Leuven, Celestijnenlaan 200D, 
B-3001 Leuven, Belgium}

\address[myquaternaryaddress]{INAF - Osservatorio Astronomico d'Abruzzo, Via mentore 
maggini s.n.c., I-64100 Teramo, Italy}

\address[myquinternaryaddress]{INFN - Sezione di Perugia, via A. Pascoli, I-06123, Perugia, 
Italy}

\address[mysecondaryaddress]{Departament de Ci\`{e}ncia de Materials i Qu\'{i}mica F\'{i}sica \& Institut de Qu\'{i}mica Te\'{o}rica i Computacional (IQTCUB), Universitat de 
Barcelona, E-08028 Barcelona, Spain}

\address[myternaryaddress]{Instituci\'{o} Catalana de Recerca i Estudis Avan\c{c}ats 
(ICREA), E-08010 Barcelona, Spain}

\begin{abstract}
We employ the Monte-Carlo Basin-Hopping (MC-BH) global optimisation technique with inter-
atomic pair potentials to generate low-energy candidates of stoichiometric alumina octomers 
((Al$_2$O$_3$)$_8$).  The candidate structures are subsequently refined with density 
functional theory calculations employing hybrid functionals (B3LYP and PBE0) and a large 
basis set (6-311+G(d)) including a vibrational analysis. We report the discovery of a set 
of energetically low-lying alumina octomer clusters, including a new global minimum 
candidate, with shapes that are elongated rather than spherical. 
%We find that the stability limit for these clusters is around a temperature of T$\simeq
%$1400 K corresponding to a phase transition in liquid alumina.
We find a stability limit for these and smaller-sized clusters at a temperature of $T\simeq1300-1450$ K corresponding to a phase transition in liquid alumina.
\end{abstract}

\begin{keyword}
aluminum oxide \sep molecular clusters \sep global optimisation \sep nucleation
%\MSC[2018] 04-26\sep  99-00
\end{keyword}

\end{frontmatter}

%\linenumbers

\section{Introduction}
Alumina clusters play a role in atmospheric chemistry 
\citep{doi:10.1029/RG020i002p00233}. Being artificially produced by rocket flights, 
Al$_2$O$_3$ cluster aerosols impact the Earth's atmospheric chemistry as they act as 
catalysts.  Moreover, owing to their high thermal stabilities and (near)-infrared 
properties, alumina clusters are promising candidates to form the seed nuclei of dust 
formation in oxygen-rich AGB stars \citep{refId0,refId0j,refId0D,Takigawaeaao2149}.
Although silicate dust constitutes the major part of oxygen-rich cosmic dust, its 
nucleation solely from gas-phase precursors is energetically hampered and explicitely ruled 
out for SiO \citep{C6CP03629E,2016A&A...591A..17G} and MgO 
\citep{doi:10.1111/j.1365-2966.2007.12358.x,1997A&A...320..553K}. 
Instead, it is more likely that the silicate dust forms on top of pre-existing seed nuclei.
These seed nuclei must form from available atoms and molecules, and have to sustain the 
extreme thermodynamic conditions close to the stellar surface. In oxygen-
dominated regimes, the latter requirement are fulfilled by highly refractory metal oxides 
such as alumina (Al$_2$O$_3$) and titania (TiO$_2$). Studies on stardust grains from pristine 
meteorites show larger sizes and greater amounts of alumina, compared with titania 
\citep{2008ApJ...682.1450N,2010ApJ...717..107G}. Thus, we presume that it is most likely 
that alumina initiates stardust synthesis in oxygen-rich conditions. \  
  
Apart from a pure scientific interest alumina clusters are also of practical use. Owing to 
their large refractories, high hardness, high stabilities, high insulation capabilities, 
and optical transparency \citep{Hart1990} alumina clusters serve as components for diverse 
purposes and applications. Over the past few decades, nano-sized alumina fibers have been 
synthesized by diverse techniques 
(hydrothermal or solvothermal, sol-gel techniques, electrospinning, extrusion, chemical 
vapor deposition) \citep{Noordin2010} and have attracted a lot of attention. 
The fibers are used in electronics, informatics, and communications \citep{Kim2014}. Moreover, 
owing to its bio-compatibility, alumina nanofibers are also used for drug delivery. In 
the laboratory, clusters of nano-sized alumina can be produced in deionized water by 
laser ablation of aluminum targets \citep{Piriyawong:2012:PCA:2489136.2489138}.

There are several bulk polymorphs for alumina. The most stable polymorph at standard 
conditions (p = 1 atm, T = 298 K) is the hexagonally, closed-packed $\alpha$-alumina which is the main component 
of corundum 
\citep{0022-3719-15-26-019}. Other alumina polymorphs are $\beta$, $\gamma$, $\delta$, $\eta
$ and $\chi$-alumina that represent the most stable alumina bulk forms at elevated 
temperatures. 
All these crystalline bulk forms have trivalent Al$^{3+}$ cations and divalent O$^{2-}$ in 
common. In this work, we compare the properties of our individual clusters with respect to 
each other, and with $\alpha$ alumina. Previous studies have shown that alumina clusters 
substantially deviate from the bulk-like analogs \citep{B212654K,doi:10.1002/anie.200604823}. In the size regime with dimension d $\le$ 1 nm (which corresponds to 
the size range from the monomer to the octomer, n=8), 
diverse geometries (flat, cage-like, compact) represent the most stable isomer structures 
\citep{doi:10.1021/jp2050614,LI2012125,ZhangCheng}. 
The alumina octomer corresponds to the first cluster size at which we find 
differences with previously reported candidate global minima in the literature. We revisit
also smaller cluster sizes (n$<$8) and we find the same structural isomers as found in the thorough study by \citep{LI2012125}.
As a result of strong (attractive and repulsive) Coulomb forces between O-anions and Al- 
cations, all favourable clusters have a strictly alternating cation-anion ordering in 
common.\\
This letter is organized as follows. In Section 2, we describe the methods to find low-energy alumina octomer structures and their subsequent refinement at the DFT (Density 
Functional Theory) level. The 
results of our calculations are presented in Section 3. 
Finally, we discuss and summarize our new findings in Section 4.

\section{Methods}
\subsection{Global optimisation searches}
\label{22}
%An alumina octomer, i.e.(Al$_{2}$O$_{3}$)$_8$, contains 40 atoms and has thus $\sum_{k=1}^{n}k$ = 40$\times$ 41/2 = 780 
%individual inter-atomic distances (that are not necessarily chemical bonds).
%All the distances between atoms contribute to the shaping of the potential energy landscape 
%resulting in a huge configurational space.
An alumina octomer, i.e.(Al$_{2}$O$_{3}$)$_8$, has 40 atoms and a total of 3$\times$40 = 120 degrees of freedom resulting in a huge configurational space.
A complete exploration of this space would require enormous amount of computational power 
and is beyond 
the current computing ability. To reduce the number of possible structural configurations 
to explore (and hence also the computational effort) a global optimisation search for low-energy octomer isomers is performed.
We employ the Monte-Carlo Basin-Hopping (MC-BH) global optimisation technique
\citep{1998cond.mat..3344W} with inter-atomic pair potentials of an Al-O system to find candidate structures. For our purposes, we use an in-house, modified version of the GMIN 
programme \citep{PhysRevLett.95.185505}.\
The general form of the inter-atomic Buckingham pair potential (including the Coulomb 
potential) reads:

\begin{equation}
U(r_{ij}) =  \frac{q_{i}q_{j}}{r} + A\exp\left(-\frac{r_{ij}}{B}\right) - \frac{C}{r_{ij}^6}
\label{buck}
\end{equation}

\noindent
where $r_{ij}$ is the relative distance of two atoms, $q_i$ and $q_j$ the charges of atom $i$ 
and $j$, respectively, and $A$, $B$ and $C$ the Buckingham pair parameters.
The first term in equation \ref{buck} represents the Coulomb law, the second term the 
short-range, steric repulsion term accounting for the Pauli principle, and the last term 
describes the attractive van-der-Waals interaction.\
The potential describes the repulsion and attraction of charged particles, in this case, 
of aluminum and oxygen ions within a (Al$_2$O$_3$)$_8$ cluster. As lightest members of 
group III and group VI elements in the periodic table,  Al and O form ionic 
bonds. The strong difference in electro-negativity, $\Delta$ EN=2.03, strongly suppresses 
the presence of covalent bonds. To reduce the probability to miss stable configurations in 
our searches we perform test calculations by swapping Al and O atoms of the most stable 
configurations. These tests account for atomic segregation (i.e. covalent bonds between 
identical atoms).  
The steric repulsion term is motivated by the fact that atoms are not dot-like but occupy a 
certain volume in space. We use the simplified form of the Buckingham pair potential and 
omit the repulsive r$^{-12}$ Lennard-Jones term. Effectively, the latter term acts only on very short distances which is already taken into account by values of the parameters 
A and B.  In the present approach we approximately account for the polarization effects by reducing the formal charges of the Al-cations and the O-anions by 10-20\%.

\begin{table}
	\caption{The parameter ranges used in this study to compute the inter-atomic Buckingham 
	pair potential. Charges q(Al) and q(O) are given in atomic units, A in eV, B in $\rm{\AA}$ C in eV $\rm{\AA}^{-6}$\label{tab1}.}	
\begin{tabular}{ l l | l l l | l l l}
	q(Al) & q(O) & A(Al-O) & B(Al-O) & C(Al-O) & A(O-O) & B(O-O) & C(O-O) \\
	\hline
	\rule{0pt}{4ex}    
	+3, +2.4 & -2, -1.6 & 2409.5 & 0.2649 & 0.0 & 22764.0 & 0.14 & 27.88 \\
	+3, +2.4 & -2, -1.6 & 4534.2 & 0.2649 & 0.0 & 25.410  & 0.6937 & 32.32\\
\end{tabular}
\end{table}

We apply two different parameter sets that are listed in Table \ref{tab1}. 
The first listed set (set 1) is commonly used for structure optimisation of Al-O systems 
(see e.g. \citep{ZhangCheng}). The second parameter set in the last line of Table 
\ref{tab1} corresponds to Ag$^{+3}$-O$^{-2}$ parameters published by \citep{A901227C}. With 
exception of the value of A(Ag-O), the other parameters are identical to the shell Al$^{3+}
$O$^{2-}$ set of \citep{JM9940400831}. 
Set 2 complements set 1 in the sense that it covers a different parameter space 
accounting for structural families that could not (or hardly) be found with set 1. For example, set 1 tends to result in compact geometries, but structural
families like void cages and open-cage-like clusters are underrepresented.   
Both parameter sets are explored with a variety of different seed structures (i.e. initial 
geometries).  
%An extensive compilation of tabulated inter-atomic parameters for a variety of elements can 
%found in the database of S. Woodley \footnote{http://www.ucl.ac.uk/klmc/Potentials/}.    

In summary, we use two different parameter sets, dozens of seed structures and 
different temperatures (T = 300-3000 K) in order 
to cover an extensive number of structural possibilities. Although the use of force fields 
is an approximation, their use enables us to perform tractable thorough searches. With our 
force-field approach we hope to have minimized the probability to miss a stable alumina 
octomer.\

\subsection{Optimisation at the DFT level}
Once multiple sets of candidate structures with different seed geometries, temperatures and 
parameter sets are found, we refine the $\sim$100 most favourable candidate structures 
with hybrid density functional theory (DFT) methods. We use two different density 
functionals, B3LYP 
\citep{1993JChPh..98.1372B} and PBE0 
\citep{1996JChPh.105.9982P}, in combination with the 6-311+G(d) basis set and perform the 
calculations with the help of the computational chemistry software package Gaussian09 
\citep{g09}. The DFT calculations are performed at 0 K and 0 atm. In the Born-Oppenheimer 
approximation used here the Potential Energy Surface (PES) does not depend on temperature.
Hence, the optimised cluster geometry is also temperature-independent. However, the 
vibrational population and the computation of the thermodynamic quantities (enthalpy, 
entropy, Gibbs free energy) depend on temperature. The thermal corrections are evaluated at 
standard conditions for temperature and pressure (T = 298.15 K, p = 1 atm). 
We include a vibrational analysis to calculate appropriate partition functions for any 
other conditions and to exclude possible transition states.

\section{Results}
\subsection{Minima structures}
\begin{figure}[!h]
\includegraphics[width=0.5\textwidth]{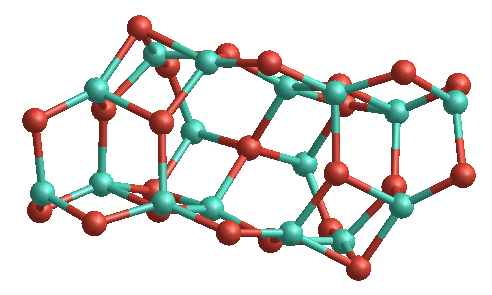}
\includegraphics[width=0.5\textwidth]{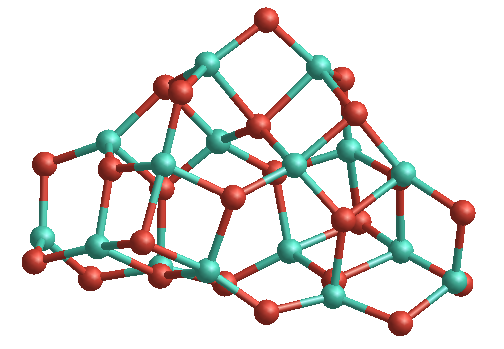}
\caption{The global minimum structure \textbf{8A} at the B3LYP/6-311+G(d) level seen from two 
distinct perspectives (that are perpendicular to each other). Al atoms are in green, O 
atoms in red.\label{c2min}}
\end{figure}

Our lowest energy candidate global minimum octomer (Al$_2$O$_3$)$_8$ structure (hereafter \textbf{8A}) calculated at the B3LYP/6-311+G(d) level of theory is displayed in Figure \ref{c2min}. The structure has a 
rotational symmetry and belongs to the C$_2$ point symmetry group. 
Its potential energy is lower by 0.15 eV relative to the next higher-lying isomer. 
With the PBE0/6-311+G(d,p) functional/basis set 8A represents the second lowest-energy 
structure (relative energy difference of 0.17 eV with respect to 8B - see below) that we find. The HOMO-LUMO gap in the B3LYP (PBE0)-
optimised new structure is 4.93 (5.45) eV.

\begin{figure}
\includegraphics[width=0.5\textwidth]{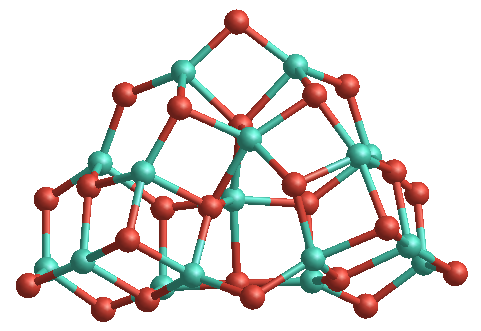}
\includegraphics[width=0.5\textwidth]{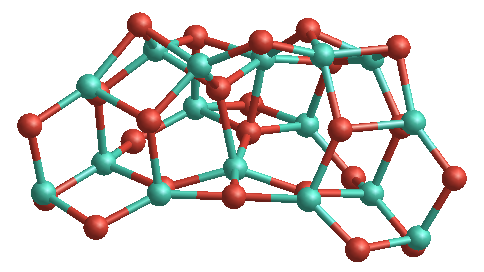}
\caption{The global minimum structure \textbf{8B} at the PBE0/6-311+G(d) level seen from two 
distinct perspectives.\label{c1min}}
\end{figure}

Our search also resulted in the discovery of another  alumina octomer cluster (hereafter 
\textbf{8B}) that is displayed in Figure \ref{c1min}. 8B shows a C$_s$-symmetric structure. 
However, during the optimisation at the DFT level, the symmetry of 8B is distorted 
and the assigned point symmetry group for 8B is C$_1$. Structure 8B is the lowest-energy configuration that we find with
the PBE0 functional and the second-lowest (relative energy 0.15 eV) with the B3LYP functional. The HOMO-LUMO gap in the B3LYP (PBE0)-optimized structure 8B is 4.67 (3.60) eV which is less than for 8A.

\begin{figure}[!h]
\captionsetup[subfigure]{labelformat=empty}
\begin{subfigure}{0.24\textwidth}
\includegraphics[width=\textwidth]{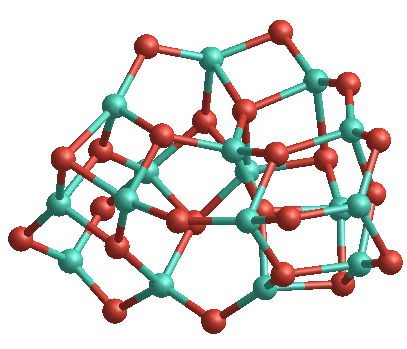}
\caption{\textbf{8C} 0.35 (0.25) eV}
\end{subfigure}
\begin{subfigure}{0.24\textwidth}
\includegraphics[width=\textwidth]{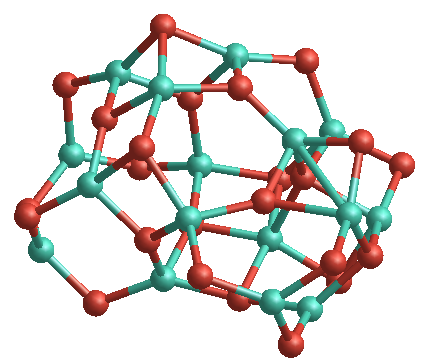}
\caption{\textbf{8D} 0.97 (1.14) eV}
\end{subfigure}
\begin{subfigure}{0.24\textwidth}
\includegraphics[width=\textwidth]{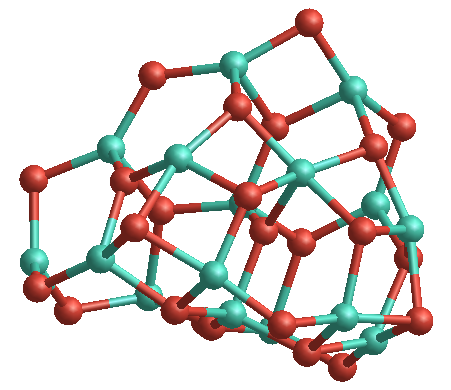}
\caption{\textbf{8E} 0.98 (0.71) eV}
\end{subfigure}
\begin{subfigure}{0.24\textwidth}
\includegraphics[width=\textwidth]{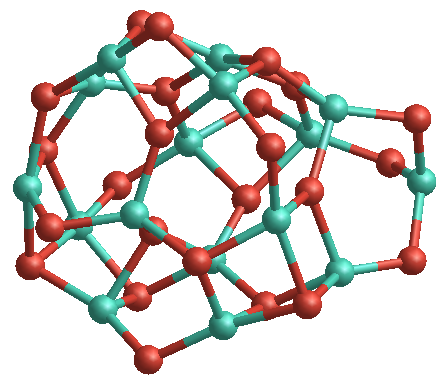}
\caption{\textbf{8F} 1.43 (1.24) eV}
\end{subfigure}
\caption{Energetically low-lying alumina octomer structures with energetic ordering according to DFT optimisations using the B3LYP functional.\label{3min}}
\end{figure}

Further energetically low-lying alumina octomer isomers that have been found in this study 
are shown in Figure \ref{3min}. All of the latter local minima structures are not symmetric 
and belong to the C$_{1}$ point group. Structure 8C has relative potential energy 0.35 eV 
(0.25 eV) above the minimum with the B3LYP (PBE0) functional. 

8D, 8E and 8F are located 0.97 (1.14), 0.98 (0.71) and 1.43 (1.24) eV above the global 
minimum structure 8A (8B), respectively. 
In particular, we note that 8D and 8E are almost degenerate at the B3LYP 
level of theory, but have a significant difference (0.43 eV) as calculated with the PBE0 
functional. All structures - except 8F, but including 8A and 8B - exhibit an overall 
elongated geometry where one dimension ($\sim$ 9-10 \AA) is significantly longer than the 
other two dimensions (typically $\sim$ 6-7 \AA). This may suggest an inherent tendency for a deviation from sphericity in a homogeneous nucleation scenario for alumina.
   
%\vspace*{1cm}
%\includegraphics[width=0.22\textwidth]{Al16O24R6L1012B.png}
%\includegraphics[width=0.20\textwidth]{Al16O24R6L1005B.png}
%\includegraphics[width=0.20\textwidth]{Al16O24R6L1009B.png}
%\includegraphics[width=0.20\textwidth]{Al16O24R7L1003B.png}

\begin{figure}[!h]
\centering
\includegraphics[width=0.3\textwidth]{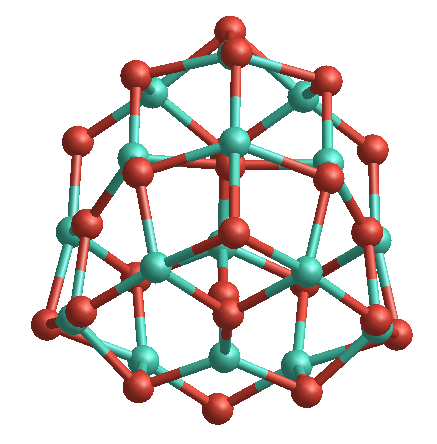}
\includegraphics[width=0.3\textwidth,angle=90]{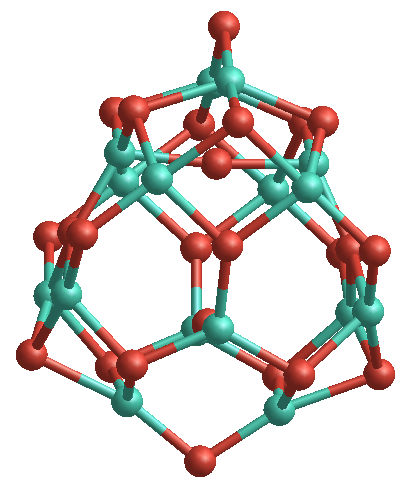}
\caption{The minimum structure \textbf{8G} seen from two 
distinct perspectives.
\label{minrah}}
\end{figure}

%\caption{.\label{5min}}
%\end{figure}
In Figure \ref{minrah}, the predicted global minimum structure in ref. \citep{doi:10.1021/jp2050614} 
(hereafter \textbf{8G}) is shown. 8G has two mirror planes, two rotational symmetries and 
belongs to the D$_{2d}$ space group. 8G has a potential energy 1.56 (0.88) eV above the 
minima structures 8A (8B).

\begin{figure}[!h]
\centering
\includegraphics[width=0.3\textwidth]{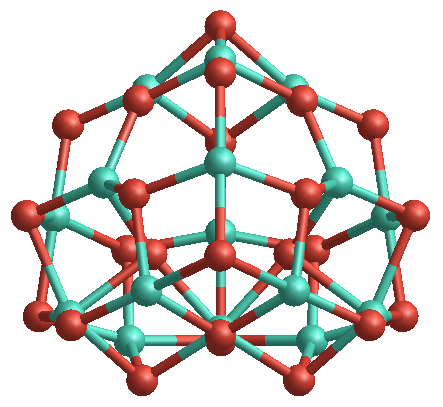}
\includegraphics[width=0.3\textwidth]{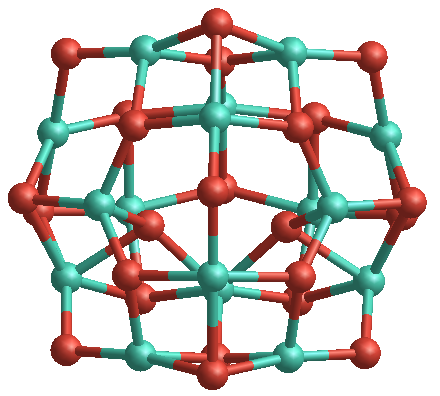}
\caption{The stable structure \textbf{8H} seen from two 
distinct perspectives.
\label{minc+e}}
\end{figure}

Five other low-lying alumina octomer isomers were predicted in ref. \citep{ZhangCheng}, see Figure 7 in their paper) using a gradient-based genetic algorithm (GA-LBFGS) 
and a Buckingham pair potential.  
We have also found the latter five isomers and investigated them on the DFT level of 
theory. Structures A, B, D and E in ref. \citep{ZhangCheng} relax to structure \textbf{8G} 
during a DFT optimisation with the hybrid functionals B3LYP and PBE0 and thus do not represent truly distinct  structural isomers. The investigation of structures C and E in Figure 7 in ref. \citep{ZhangCheng} at a DFT level results in the same isomer (see Figure \ref{minc+e}) with a relative energy 2.77 (1.89) eV above 8A (8B). It has a mirror plane and space group C$_{2v}$.

\begin{figure}[!h]
\centering
\includegraphics[width=0.3\textwidth]{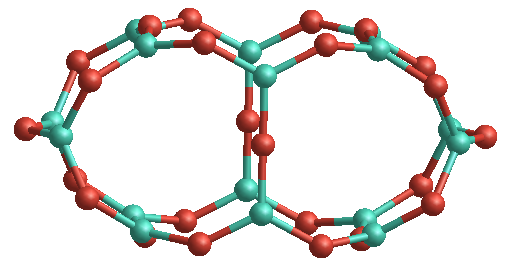}
\includegraphics[width=0.3\textwidth]{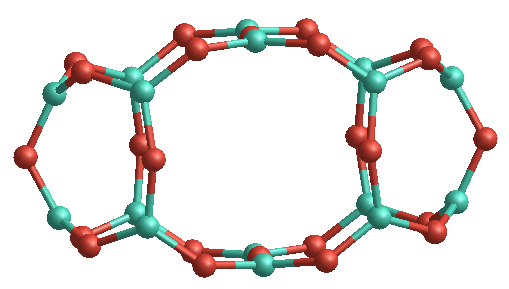}
\caption{The stable structure \textbf{8I} 
\label{minpeanut}}
\end{figure}

Recently, the structural family of (Al$_2$O$_3$)$_n$ hollow spheres and bubble clusters has 
been studied for certain sizes (n=10, 12, 16, 18, 24 and 33) \citep{GU201529}. The latter 
structures 
are characterized by Al-atoms with coordination number 3 and O-atoms that are 2-
coordinated. For n=8, we also find a member of the hollow spheres, or ``bubble'' family 
(see Figure \ref{minpeanut}). Cluster 8I has a very symmetric peanut-shaped geometry and 
belongs to the D$_{2h}$ point group. It has an electronic energy 5.31 (7.67) eV above the 
global minimum.\\
We report the finding of six new isomer structures (8A, 8B, 8C, 8D, 8E, 8F).  
The previously reported structures 8G, 8H and 8I have potential 
energies that are significantly higher than 8A-8F.  

\subsection{Rotational constants}
The rotational constants of the two isomers 8A and 8B are tabulated in Table 
\ref{rotc}. 
The rotational constants obtained with the PBE0 functional are marginally larger 
($\sim$ 1\%) that those obtained with the B3LYP functional. This is a consequence 
of the slightly more compact geometry of the PBE0 optimized structures leading 
to lower moments of inertia and higher rotational constants. Owing to its C$_2$ 
symmetry 8B has two almost identical constants (B$_y$ and B$_z$).

\begin{table}
\caption{Rotational constants for 8A and 8B in (GHz)\label{rotc}}
\begin{tabular}{l | r r | r r}
Rot. Constants & \multicolumn{2}{l}{8A} & \multicolumn{2}{l}{8B} \\
GHz & (B3LYP) & PBE0 &  B3LYP & PBE0 \\
\hline
B$_x$ & 0.113953 & 0.115163 & 0.090181 & 0.091319 \\
B$_y$ & 0.062404 & 0.063075 & 0.084215 & 0.085045 \\
B$_z$ & 0.056707 & 0.057285 & 0.084212 & 0.085042 \\
\end{tabular}
\end{table}

\subsection{Bond distances, coordination and charges}
\begin{figure}[!h]
\includegraphics[width=\textwidth]{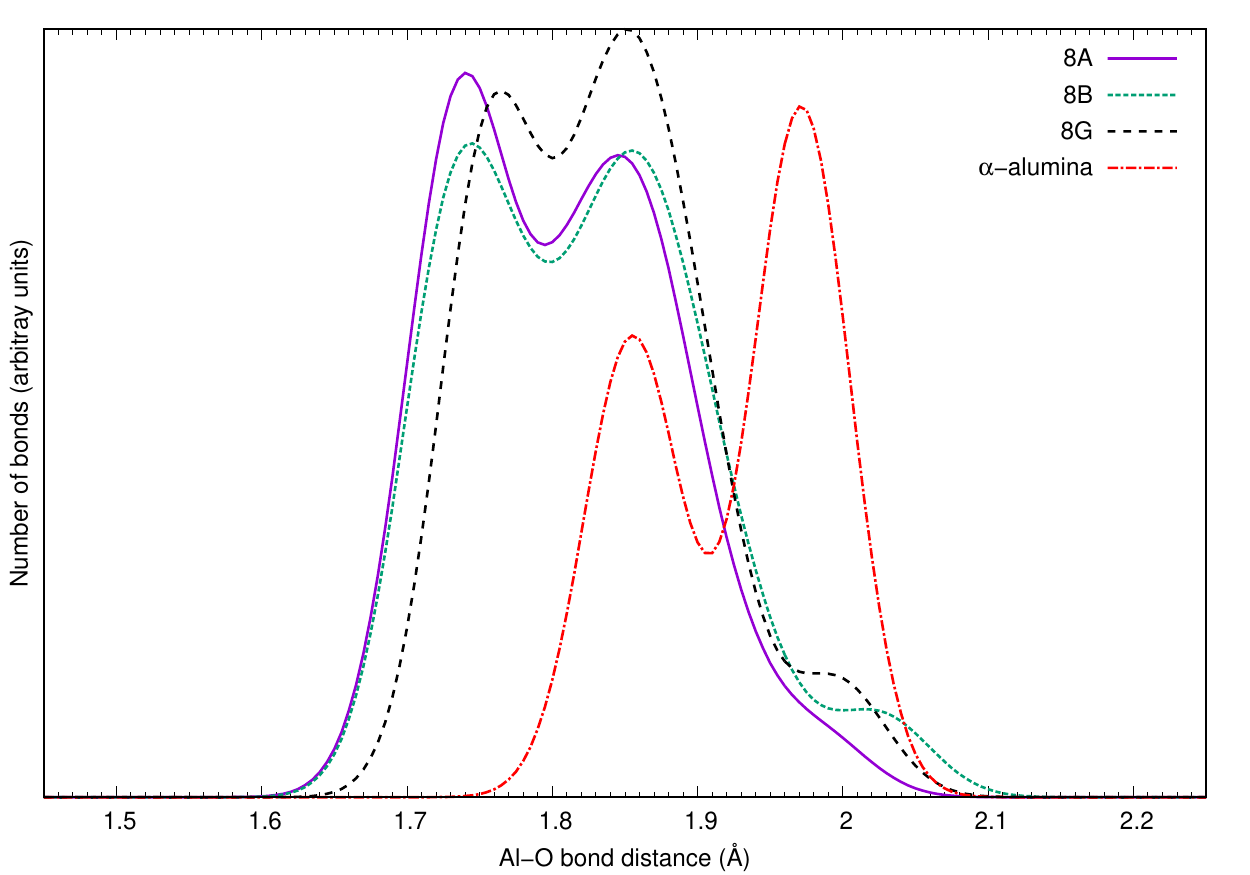}
\caption{The Al-O bond lengths of 8A, 8B, 8G and $\alpha$-alumina. The curves are fitted with a Gaussian distribution and a half width $\gamma$=0.03$\rm{\AA}$ \label{bonddist}.}
\end{figure}

The Al-O bond distances of the clusters 8A, 8B, and 8G, and of the crystalline bulk ($
\alpha$-alumina) are shown in Figure \ref{bonddist}. Al-Al and O-O bonds do not 
appear owing to the strong Coulomb repulsion of ions with the same (or similar) charge. 
The Al-O bond lengths of the clusters 8A, 8B, and 8G, as well as of $\alpha$-alumina 
form two populations, at short and at long bond lengths. 
The short bond peaks at 1.74 $\rm{\AA}$, and the long bond at 1.845 for structure 8A,   
for 8B the peaks are located at 1.745 $\rm{\AA}$ and 1.855 $\rm{\AA}$, whereas they lie 
around 1.765 $\rm{\AA}$ and 1.850 $\rm{\AA}$ for 8G. Also in the bulk phase, $\alpha$ 
alumina exhibits Al-O bonds with two different lengths being located at 1.972 $\rm{\AA}$ and at 1.855 $\rm{\AA}$.

It is apparent that the clusters and the bulk exhibit bond lengths of around 1.85 $\rm{\AA}$. 
However, the most prominent feature of $\alpha$-alumina located at 1.972 $\rm{\AA}$ is largely absent 
in the clusters and vice versa,       
inter-atomic distances smaller than 1.8 $\rm{\AA}$ do no appear in $\alpha$-alumina, but account 
for a significant fraction of the cluster bonds.\
In 8A, 14 out of 16 Al cations are 4-coordinated, the remaining 2 Al are 3-coordinated, 
whereas 11 oxygen anions are 2-coordinated, 12 O atoms 3-coordinated and just one 4-
coordinated. We count 62 Al-O bonds in total.\
The situation is similar for isomer 8B: It has 12 4-coordinated, 2 3-coordinated and 2 5-
coordinated Al cations, respectively. The oxygen anions are 2-coordinated (10), 3-
coordinated (12) and 4-coordinated (2). A total of 64 Al-O bonds is present in 8B.\
Contrary, structure 8G exhibits 12 4-coordinated, 4 5-coordinated, but no 3-coordinated Al 
cations. Also in the oxygen coordination 8G differs from 8A and 8B: 16 O anions are 3 
coordinated, 6 O ions are 2-coordinated and 2 O ions are 4-coordinated, respectively. The 
number of 
Al-O bonds is 68 and thus slightly higher than in 8A and 8B.\
In $\alpha$-alumina, Al cations are 6-coordinated and O anion 4-coordinated. The average 
coordination in the crystalline bulk is (as expected) higher than for the clusters. In the 
clusters, a considerable fraction of the atoms are located on the surface. Consequently, 
part of their atomic "neighbors" are missing and their coordination is lower.\ 

In Table \ref{mullik}, the Mulliken charges of the presented clusters are shown. 
We also include a value for the average charge of $\alpha$-alumina in the bulk phase (q($
\overline{Al}$)=+1.38 e, \citep{Pinto2004}). However, we note that formal charge 
calculations strongly depend on the used basis set and functional and a comparison to the clusters is biased. 
In general, we find that the most stable isomers have higher formal atomic charges than
the energetically less favourable clusters.
Cluster 8I represents an exception, as it exhibits the largest 
average charges of all investigated clusters, but represents one of the energetically least 
stable ones due to its comparatively low coordination.

\begin{table}
\caption{Mulliken charge analysis for isomers 8A, 8B, 8G and $\alpha$-alumina (in atomic 
units)\label{mullik}}
\begin{tabular}{l | c c c c | c}
charge & 8A & 8B & 8G & 8I & $\alpha$-alumina \\
\hline
\rule{0pt}{4ex}q($\overline{Al}$)  &  0.42 (0.30)  &  0.34 (0.24) & 0.22 (0.09) & 0.46 (0.36) & 1.38 \citep{Pinto2004} \\
q($\overline{O}$)   & -0.28 (-0.20) & -0.23 -(0.16) & -0.15 (-0.06) & -0.31 (-0.23) & -0.92(=2/3$\times$1.38) \\
\hline
%q$_{max}$(Al) & 0.96 &  0.86 & 0.88 &  \\
%q$_{min}$(Al) & -0.47 & -0.53 & -1.13 & \\
%q$_{max}$(O)  & 0.33  & 0.27  & 0.77 & \\
%q$_{min}$(O)  & -0.50 & -0.54 & -0.69 & \\
\end{tabular}
\end{table}

\subsection{Bond angles}
\begin{figure}[!h]
\includegraphics[width=\textwidth]{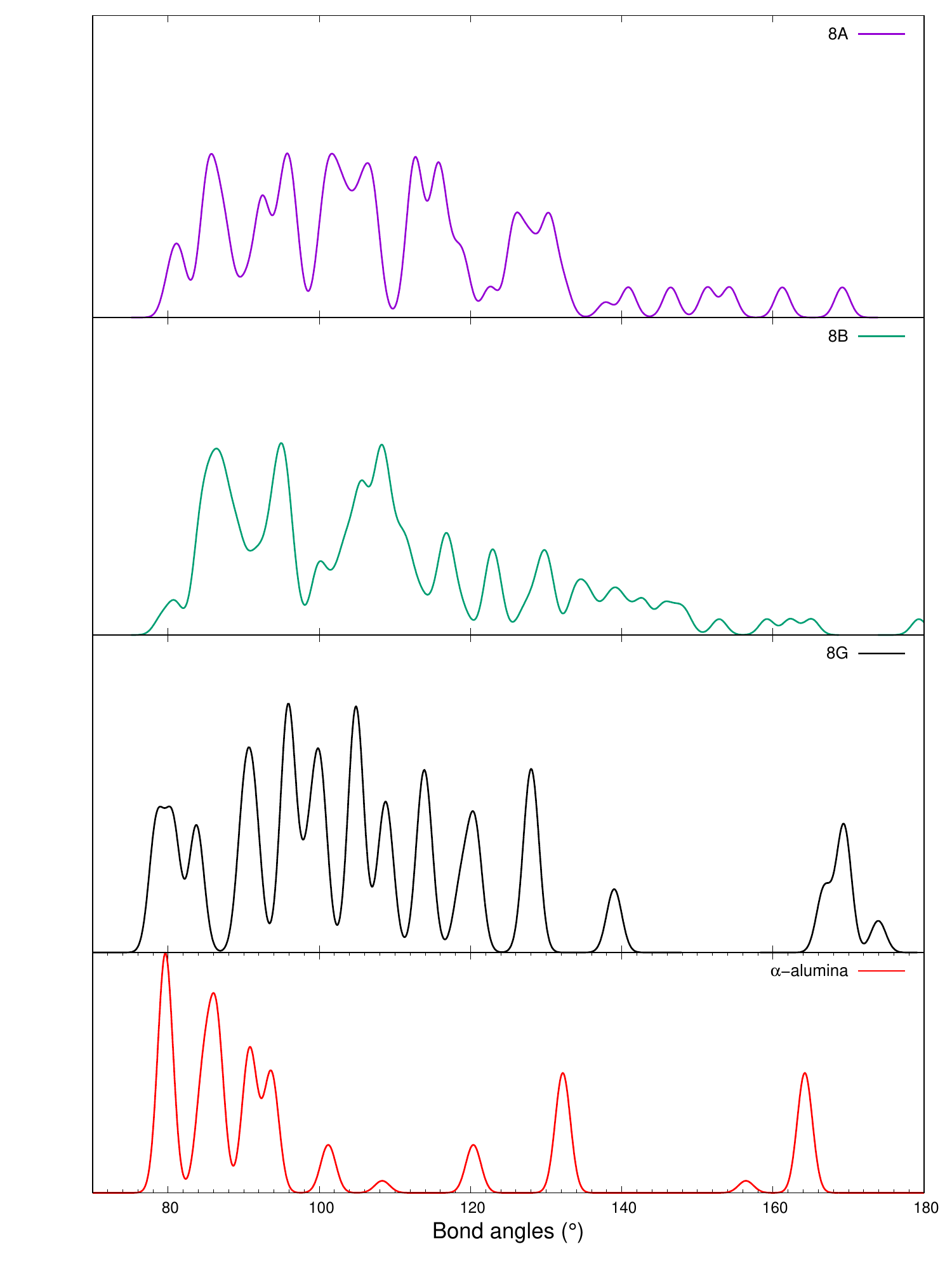}
\caption{The Al-O-Al and O-Al-O bond angles of structures 8A 8B, 8G and alpha-alumina. The curves are fitted with a Gaussian distribution and a half width $\gamma$=1.0$^{\circ}$ 
\label{angles}.}
\end{figure}

In Figure \ref{angles}, the Al-O-Al and O-Al-O bond angles are displayed for structures 8A, 8B, 8G 
and $\alpha$-alumina. The latter exhibits characteristic angles at 79.6$^{\circ}$, 84.6$^{\circ}$, 86.4$^{\circ}$, 
90.8$^{\circ}$, 93.6$^{\circ}$ and 132.2$^{\circ}$ and 164.2$^{\circ}$ degrees. The isomers 8A and 8B show a broad distribution of 
bond angles in the range between 80 and 180 degrees. Compared with 8G, and the crystalline 
phase ($\alpha$-alumina) the angle distribution of 8A and 8B is broader and has less 
pronounced peaks. The latter can be explained by the higher degree of symmetry in structure 
8G, since owing to its internal plane symmetry, every angle appears twice. The crystalline 
form of alumina shows the fewest and most pronounced peaks, owing to symmetry reasons.
    
The difference of the bulk to the clusters can be explained by the finite-size geometries of 
the clusters. 
Whereas in $\alpha$-alumina the periodicity of the 
crystal implies a homogeneous spatial occupation by the Al and O atoms, the clusters are 
largely empty in their interiors and the atoms reside on the surface. As a consequence the 
cluster bond angles are systematically larger than in $\alpha$-alumina.

\subsection{Vibrational analysis}
\begin{figure}[h!]
\includegraphics[width=\textwidth]{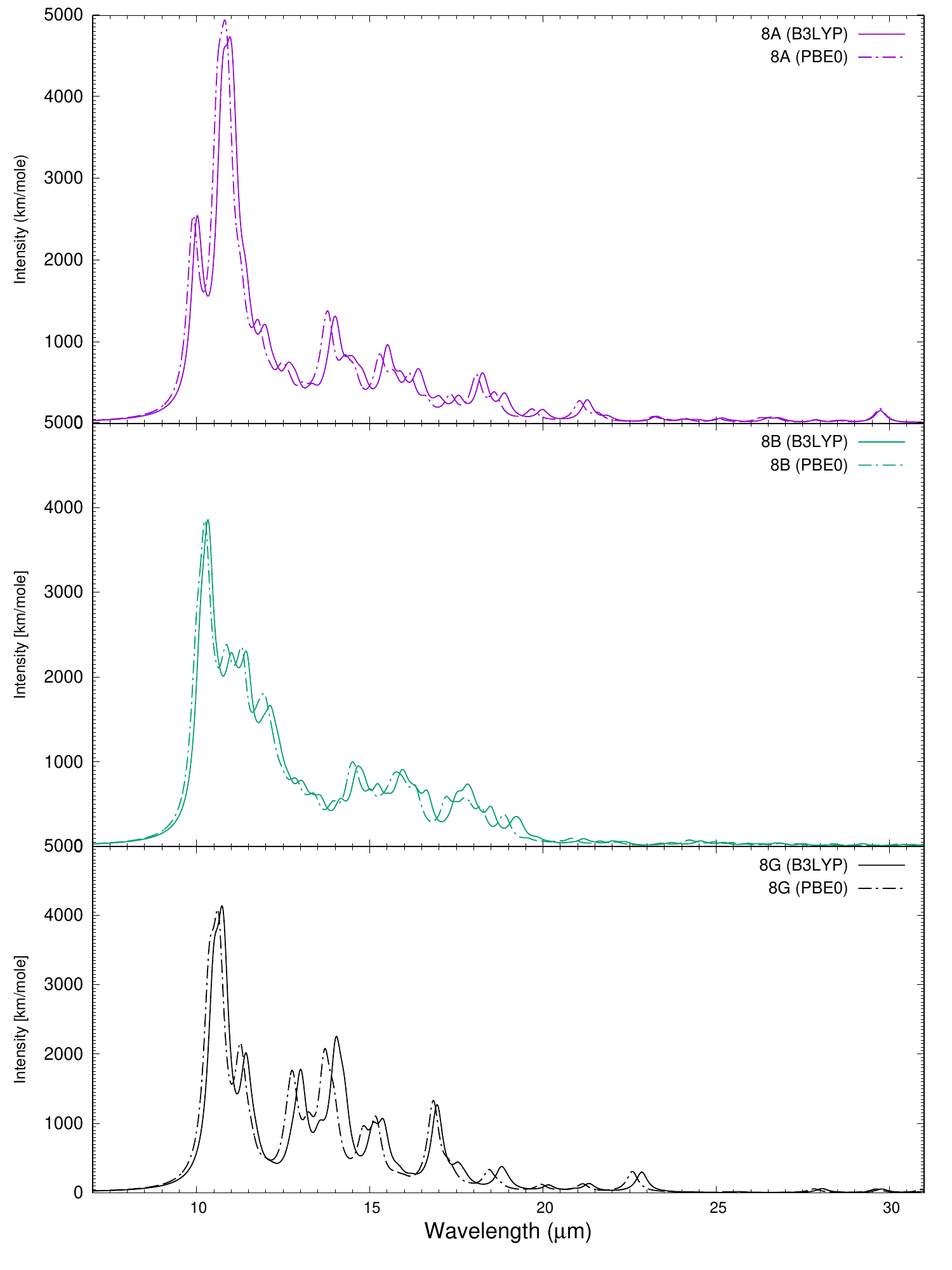}
\caption{The vibration modes and related intensities of the newly-discovered clusters 8A, 8B, and 8G fitted by a Lorentzian profile with a half width $\gamma$=0.2. Straight lines: B3LYP, dashed lines: PBE0\label{vibs}.}
\end{figure}

An alumina octomer has 40 atoms, consequently the number of vibrational degrees of 
freedom (i.e. vibration modes) is $3\times40-6 = 114$.
The vibrational infrared (IR) spectra of 8A, 8B, and 8G are shown as a function of  
wavelength in Figure \ref{vibs}. We use a Lorentzian function to describe the 
distribution of the peaks with a half-width at half-maximum $\gamma$= 0.2.

The majority of the IR modes of all shown clusters are located in a wavelength range 
between 10 and 20 $\mu$m with a culmination point at 10-11 $\mu$m. For structure 8A, the most intense vibration modes are located at 
wavelengths 10.012 (9.894), 10.985 (10.828), 10.729 (10.587) and 11.064 (10.921) $\mu$m for the B3LYP (PBE0) functional, respectively.
For cluster 8B, the 10.348 (10.253), 10.126 (10.005), 11.009 (10.871) and 11.473 (11.329) $
\mu$m for the B3LYP (PBE0) functional, 
respectively. These latter dominating vibrations are attributed to Al-O
stretching and bending modes. A complete table with all vibrational frequencies for 8A and 8B can be found in the Appendix.

We investigate isomers 8A, 8B, and 8G also with the SDD (Stuttgart/Dresden) basis set 
including a vibrational analysis in order to benchmark with the results of ref. \citep{doi:10.1021/jp2050614} 
who predicted isomer 8G as the global minimum.  However, we found that vibrational calculations employing the SDD basis set led to imaginary 
frequencies for 8A, 8B, 8G which do not appear using a larger basis set (6-311+G(p)). 
We conclude that the SDD basis set is not adequate to describe alumina clusters  
and that all vibrational frequencies are real. Hence, 
clusters 8A, 8B, and 8G correspond to real minimum structures (and not transition states).
 
The differences in the IR spectra between  PBE0- and B3LYP-based calculations are small and the tiny relative shifts in wavelengths arise - among other reasons - due to the slightly shorter bond distances obtained with the PBE0 functional, compared to the B3LYP functional (as for the rotational constants).\

The vibrational zero-point energies of isomers 8A, 8B, and 8G are given in Table \ref{zero}.
They are of the order 3.5 eV and vary by only 0.02 eV among the cluster isomers 8A, 8B, and 8G within the same level of theory. Consequently, the relative energies of the clusters hardly shift and the energetic ordering is preserved by including the vibrational zero-point correction.  

\begin{table}
\caption{Vibrational zero-point corrections $\epsilon_0$ (in eV) of 8A, 8B, and 8G as 
calculated with the B3LYP and the PBE0 functional\label{zero}.}
\center
\begin{tabular}{l | r r }
Isomer & $\epsilon_0$(B3LYP) & $\epsilon_0$(PBE0)\\
\hline
8A & 3.50 & 3.54 \\
8B & 3.51 & 3.55 \\
8G & 3.49 & 3.54 \\
\end{tabular}
\end{table}

The vibration frequencies of the clusters 8A, 8B, and 8G are shown in units of 
wavenumbers (cm$^{-1}$) in Table \ref{frequen} in the Appendix. For clusters 8A and 8B, 
the three lowest modes are located slightly below 130 cm$^{-1}$. 
In the harmonic oscillator approximation used in the present study, these low frequency modes are associated with hindered rotations rather than vibrations. No imaginary frequency occurs for structures  8A, 8B, and 8G.\ 

\subsection{Thermochemistry}
We calculate thermochemical potentials (heat capacity c$_P$, entropy S, enthalpy of 
formation $\Delta$H$_{f}$(T) and the (Gibbs) free energy of formation) $\Delta$G$_{f}$(T) 
of both clusters 8A and 8B as a function of temperature (see Tables \ref{td8a} and \ref{td8b} in the Appendix). We use a 
script provided by \citep{doi:10.1021/bk-1998-0677} to calculate the heat capacity c$_p$(T), entropy S(T) and enthalpy H$^0$(T) from the cluster calculation output.
In order to determine dH$_f$ and dG$_f$, we follow the approach of Ochterski (see \citep{g09}) and 
make use of the enthalpies and the entropies of the elements (Al and O) in their standard 
states listed the NIST-JANAF Thermochemical Tables \footnote{https://janaf.nist.gov/}. As the fugacity of elemental gas-phase Al is 1 bar at 2790.81 K we constrain the temperature range to a maximum of T = 2700 K. 

The thermodynamic tables for structures 8A and 8B can be found in the Appendix.
They are evaluated at a pressure of p = 1 atm. 
At 0 K the entropy dS is vanishing by definition and dH$_f$ equals dG$_f$. We derive dH$_f$ = -9114.5 kJ/mole for 8A and dH$_f$ = -9100.3 kJ/mole 
for 8B $-$ in agreement with the relative potential energies including the zero-point correction.
At 298.15 K the situation is similar and the relative enthalpies and Gibbs free energies differ by
13$-$17 kJ mole$^{-1}$. Also for larger temperatures, structure 8A is slightly more favored to form than structure 8B as calculated within the B3lYP density functional.   
Around T $=$ 1400$-$1500 K, the Gibbs free energy of formation for structures 8A and 8B changes sign and becomes positive (endergonic) for larger temperatures. 
The stability limits of the most stable, smaller-sized (n$<8$) alumina clusters are similar to those of 8A and 8B, and range from T=1300-1450 K. We find a slight increase of the critical temperature with cluster size.
Thus, an effective alumina 
octomer formation is expected to occur for temperatures below 1400 K and to be hampered for temperatures larger than 1500 K (at standard pressure of 1 atm). Interestingly, the latter 
temperature coincidences with the condensation temperature for alumina grains in circumstellar conditions\citep{2010LNP...815..143M}.\\

\section{Conclusion} 
We report the discovery of six (Al$_2$O$_3$)$_8$ isomers (8A, 8B, 8C, 8D, 8E, 8F) that represent currently the lowest-energy alumina octomer structures known. 
The octomer is the smallest cluster size at which we find differences with previously reported candidate global minima.
All six favourable octomer structures are neither  
hollow nor spherical, but exhibit elongated and ``semi-compact'' geometries. The IR spectra
of the lowest-energy octomers is dominated by Al-O vibration modes in 10-11 $\mu$m 
wavelength range. By comparing formal charges, bond lengths and angles of the reported low-
energy 
octomers to those of crystalline $\alpha$-alumina, we conclude that, at a cluster size n=8, 
the alumina 
bulk limit is not (yet) reached. 
However, we find that the thermal stability limit of the 
octomers and smaller-sized polymers in the gas phase coincides with the glass-
liquid phase transition of liquid alumina.\\

%\section{Acknowledgements}
\textit{Acknowledgements}\ 
\textit{This work was supported by the ERC consolidator grant 646758 ``AEROSOL''.
We acknowledge Dr. Amol D. Rahane for providing the cluster geometries from their studies for comparison. 
Moreover, we acknowledge the ``Accordo Quadro INAF-CINECA (2017)'', for the
availability of high performance computing resources and support. This work was supported by a STSM Grant from COST Action CM1401 (``Our AstroChemical History'')}.
This research was supported by the Spanish MINECO/FEDER CTQ2015-64618-R grant and, in part, by Generalitat de Catalunya (grants 2017SGR13 and XRQTC).

\section*{References}

\bibliography{biblio}

\begin{thebibliography}{10}
\expandafter\ifx\csname url\endcsname\relax
  \def\url#1{\texttt{#1}}\fi
\expandafter\ifx\csname urlprefix\endcsname\relax\def\urlprefix{URL }\fi
\expandafter\ifx\csname href\endcsname\relax
  \def\href#1#2{#2} \def\path#1{#1}\fi

\bibitem{doi:10.1029/RG020i002p00233}
R.~P. Turco, R.~C. Whitten, O.~B. Toon,
  \href{https://agupubs.onlinelibrary.wiley.com/doi/abs/10.1029/RG020i002p00233}{Stratospheric
  aerosols: Observation and theory}, Reviews of Geophysics 20~(2) (1982)
  233--279.
\newblock \href
  {http://arxiv.org/abs/https://agupubs.onlinelibrary.wiley.com/doi/pdf/10.1029/RG020i002p00233}
  {\path{arXiv:https://agupubs.onlinelibrary.wiley.com/doi/pdf/10.1029/RG020i002p00233}},
  \href {http://dx.doi.org/10.1029/RG020i002p00233}
  {\path{doi:10.1029/RG020i002p00233}}.
\newline\urlprefix\url{https://agupubs.onlinelibrary.wiley.com/doi/abs/10.1029/RG020i002p00233}

\bibitem{refId0}
{Karovicova, I.}, {Wittkowski, M.}, {Ohnaka, K.}, {Boboltz, D. A.}, {Fossat,
  E.}, {Scholz, M.}, \href{https://doi.org/10.1051/0004-6361/201322376}{New
  insights into the dust formation of oxygen-rich agb stars}, A\&A 560 (2013)
  A75.
\newblock \href {http://dx.doi.org/10.1051/0004-6361/201322376}
  {\path{doi:10.1051/0004-6361/201322376}}.
\newline\urlprefix\url{https://doi.org/10.1051/0004-6361/201322376}

\bibitem{refId0j}
{Gobrecht, D.}, {Cherchneff, I.}, {Sarangi, A.}, {Plane, J. M. C.}, {Bromley,
  S. T.}, \href{https://doi.org/10.1051/0004-6361/201425363}{Dust formation in
  the oxygen-rich agb star ik tauri}, A\&A 585 (2016) A6.
\newblock \href {http://dx.doi.org/10.1051/0004-6361/201425363}
  {\path{doi:10.1051/0004-6361/201425363}}.
\newline\urlprefix\url{https://doi.org/10.1051/0004-6361/201425363}

\bibitem{refId0D}
{Decin, L.}, {Richards, A. M. S.}, {Waters, L. B. F. M.}, {Danilovich, T.},
  {Gobrecht, D.}, {Khouri, T.}, {Homan, W.}, {Bakker, J. M.}, {Van de Sande,
  M.}, {Nuth, J. A.}, {De Beck, E.},
  \href{https://doi.org/10.1051/0004-6361/201730782}{Study of the aluminium
  content in agb winds using alma - indications for the presence of gas-phase
  (al2o3)n clusters}, A\&A 608 (2017) A55.
\newblock \href {http://dx.doi.org/10.1051/0004-6361/201730782}
  {\path{doi:10.1051/0004-6361/201730782}}.
\newline\urlprefix\url{https://doi.org/10.1051/0004-6361/201730782}

\bibitem{Takigawaeaao2149}
A.~Takigawa, T.~Kamizuka, S.~Tachibana, I.~Yamamura,
  \href{http://advances.sciencemag.org/content/3/11/eaao2149}{Dust formation
  and wind acceleration around the aluminum oxide-rich agb star w hydrae},
  Science Advances 3~(11).
\newblock \href
  {http://arxiv.org/abs/http://advances.sciencemag.org/content/3/11/eaao2149.full.pdf}
  {\path{arXiv:http://advances.sciencemag.org/content/3/11/eaao2149.full.pdf}},
  \href {http://dx.doi.org/10.1126/sciadv.aao2149}
  {\path{doi:10.1126/sciadv.aao2149}}.
\newline\urlprefix\url{http://advances.sciencemag.org/content/3/11/eaao2149}

\bibitem{C6CP03629E}
S.~T. Bromley, J.~C. Gomez~Martin, J.~M.~C. Plane,
  \href{http://dx.doi.org/10.1039/C6CP03629E}{Under what conditions does (sio)n
  nucleation occur? a bottom-up kinetic modelling evaluation}, Phys. Chem.
  Chem. Phys. 18 (2016) 26913--26922.
\newblock \href {http://dx.doi.org/10.1039/C6CP03629E}
  {\path{doi:10.1039/C6CP03629E}}.
\newline\urlprefix\url{http://dx.doi.org/10.1039/C6CP03629E}

\bibitem{2016A&A...591A..17G}
H.-P. {Gail}, M.~{Scholz}, A.~{Pucci}, {Silicate condensation in Mira
  variables}, aap 591 (2016) A17.
\newblock \href {http://arxiv.org/abs/1604.04636} {\path{arXiv:1604.04636}},
  \href {http://dx.doi.org/10.1051/0004-6361/201628113}
  {\path{doi:10.1051/0004-6361/201628113}}.

\bibitem{doi:10.1111/j.1365-2966.2007.12358.x}
J.~S. Bhatt, I.~J. Ford,
  \href{http://dx.doi.org/10.1111/j.1365-2966.2007.12358.x}{Investigation of
  mgo as a candidate for the primary nucleating dust species around m stars},
  Monthly Notices of the Royal Astronomical Society 382~(1) (2007) 291--298.
\newblock \href
  {http://arxiv.org/abs//oup/backfile/content_public/journal/mnras/382/1/10.1111/j.1365-2966.2007.12358.x/2/mnras0382-0291.pdf}
  {\path{arXiv:/oup/backfile/content_public/journal/mnras/382/1/10.1111/j.1365-2966.2007.12358.x/2/mnras0382-0291.pdf}},
  \href {http://dx.doi.org/10.1111/j.1365-2966.2007.12358.x}
  {\path{doi:10.1111/j.1365-2966.2007.12358.x}}.
\newline\urlprefix\url{http://dx.doi.org/10.1111/j.1365-2966.2007.12358.x}

\bibitem{1997A&A...320..553K}
T.~M. {Koehler}, H.-P. {Gail}, E.~{Sedlmayr}, {MgO dust nucleation in M-Stars:
  calculation of cluster properties and nucleation rates.}, aap 320 (1997)
  553--567.

\bibitem{2008ApJ...682.1450N}
L.~R. {Nittler}, C.~M.~O. {Alexander}, R.~{Gallino}, P.~{Hoppe}, A.~N.
  {Nguyen}, F.~J. {Stadermann}, E.~K. {Zinner}, {Aluminum-, Calcium- and
  Titanium-rich Oxide Stardust in Ordinary Chondrite Meteorites}, apj 682
  (2008) 1450--1478.
\newblock \href {http://arxiv.org/abs/0804.2866} {\path{arXiv:0804.2866}},
  \href {http://dx.doi.org/10.1086/589430} {\path{doi:10.1086/589430}}.

\bibitem{2010ApJ...717..107G}
F.~{Gyngard}, E.~{Zinner}, L.~R. {Nittler}, A.~{Morgand}, F.~J. {Stadermann},
  K.~{Mairin Hynes}, {Automated NanoSIMS Measurements of Spinel Stardust from
  the Murray Meteorite}, apj 717 (2010) 107--120.
\newblock \href {http://arxiv.org/abs/1006.4355} {\path{arXiv:1006.4355}},
  \href {http://dx.doi.org/10.1088/0004-637X/717/1/107}
  {\path{doi:10.1088/0004-637X/717/1/107}}.

\bibitem{Hart1990}
L.~D. Hart, \href{http://www.osti.gov/scitech/servlets/purl/5158074}{Alumina
  chemicals}, Columbus, OH (USA); American Ceramic Society Inc., United States,
  1990.
\newline\urlprefix\url{http://www.osti.gov/scitech/servlets/purl/5158074}

\bibitem{Noordin2010}
M.~R. Noordin, K.~Y. Liew, \href{http://dx.doi.org/10.5772/8165}{Synthesis of
  Alumina Nanofibers and Composites}, InTech, Rijeka, 2010.
\newblock \href {http://dx.doi.org/10.5772/8165} {\path{doi:10.5772/8165}}.
\newline\urlprefix\url{http://dx.doi.org/10.5772/8165}

\bibitem{Kim2014}
J.-H. Kim, S.-J. Yoo, D.-H. Kwak, H.-J. Jung, T.-Y. Kim, K.-H. Park, J.-W. Lee,
  \href{https://doi.org/10.1186/1556-276X-9-44}{Characterization and
  application of electrospun alumina nanofibers}, Nanoscale Research Letters
  9~(1) (2014) 44.
\newblock \href {http://dx.doi.org/10.1186/1556-276X-9-44}
  {\path{doi:10.1186/1556-276X-9-44}}.
\newline\urlprefix\url{https://doi.org/10.1186/1556-276X-9-44}

\bibitem{Piriyawong:2012:PCA:2489136.2489138}
V.~Piriyawong, V.~Thongpool, P.~Asanithi, P.~Limsuwan,
  \href{http://dx.doi.org/10.1155/2012/819403}{Preparation and characterization
  of alumina nanoparticles in deionized water using laser ablation technique},
  J. Nanomaterials 2012 (2012) 2:2--2:2.
\newblock \href {http://dx.doi.org/10.1155/2012/819403}
  {\path{doi:10.1155/2012/819403}}.
\newline\urlprefix\url{http://dx.doi.org/10.1155/2012/819403}

\bibitem{0022-3719-15-26-019}
I.~P. Batra, \href{http://stacks.iop.org/0022-3719/15/i=26/a=019}{Electronic
  structure of α-al 2 o 3}, Journal of Physics C: Solid State Physics 15~(26)
  (1982) 5399.
\newline\urlprefix\url{http://stacks.iop.org/0022-3719/15/i=26/a=019}

\bibitem{B212654K}
D.~van Heijnsbergen, K.~Demyk, M.~A. Duncan, G.~Meijer, G.~von Helden,
  \href{http://dx.doi.org/10.1039/B212654K}{Structure determination of gas
  phase aluminum oxide clusters}, Phys. Chem. Chem. Phys. 5 (2003) 2515--2519.
\newblock \href {http://dx.doi.org/10.1039/B212654K}
  {\path{doi:10.1039/B212654K}}.
\newline\urlprefix\url{http://dx.doi.org/10.1039/B212654K}

\bibitem{doi:10.1002/anie.200604823}
M.~Sierka, J.~Döbler, J.~Sauer, G.~Santambrogio, M.~Brümmer, L.~Wöste,
  E.~Janssens, G.~Meijer, K.~Asmis,
  \href{https://onlinelibrary.wiley.com/doi/abs/10.1002/anie.200604823}{Unexpected
  structures of aluminum oxide clusters in the gas phase}, Angewandte Chemie
  International Edition 46~(18) (2007) 3372--3375.
\newblock \href
  {http://arxiv.org/abs/https://onlinelibrary.wiley.com/doi/pdf/10.1002/anie.200604823}
  {\path{arXiv:https://onlinelibrary.wiley.com/doi/pdf/10.1002/anie.200604823}},
  \href {http://dx.doi.org/10.1002/anie.200604823}
  {\path{doi:10.1002/anie.200604823}}.
\newline\urlprefix\url{https://onlinelibrary.wiley.com/doi/abs/10.1002/anie.200604823}

\bibitem{doi:10.1021/jp2050614}
A.~B. Rahane, M.~D. Deshpande, V.~Kumar,
  \href{https://doi.org/10.1021/jp2050614}{Structural and electronic properties
  of (al2o3)n clusters with n = 1–10 from first principles calculations}, The
  Journal of Physical Chemistry C 115~(37) (2011) 18111--18121.
\newblock \href {http://arxiv.org/abs/https://doi.org/10.1021/jp2050614}
  {\path{arXiv:https://doi.org/10.1021/jp2050614}}, \href
  {http://dx.doi.org/10.1021/jp2050614} {\path{doi:10.1021/jp2050614}}.
\newline\urlprefix\url{https://doi.org/10.1021/jp2050614}

\bibitem{LI2012125}
R.~Li, L.~Cheng,
  \href{http://www.sciencedirect.com/science/article/pii/S2210271X12003726}{Structural
  determination of (al2o3)n (n=1–7) clusters based on density functional
  calculation}, Computational and Theoretical Chemistry 996 (2012) 125 -- 131.
\newblock \href
  {http://dx.doi.org/https://doi.org/10.1016/j.comptc.2012.07.027}
  {\path{doi:https://doi.org/10.1016/j.comptc.2012.07.027}}.
\newline\urlprefix\url{http://www.sciencedirect.com/science/article/pii/S2210271X12003726}

\bibitem{ZhangCheng}
Q.~Zhang, L.~Cheng, \href{https://doi.org/10.1021/acs.jcim.5b00069}{Structural
  determination of (al2o3)n (n = 1–15) clusters based on graphic processing
  unit}, Journal of Chemical Information and Modeling 55~(5) (2015) 1012--1020,
  pMID: 25928795.
\newblock \href {http://arxiv.org/abs/https://doi.org/10.1021/acs.jcim.5b00069}
  {\path{arXiv:https://doi.org/10.1021/acs.jcim.5b00069}}, \href
  {http://dx.doi.org/10.1021/acs.jcim.5b00069}
  {\path{doi:10.1021/acs.jcim.5b00069}}.
\newline\urlprefix\url{https://doi.org/10.1021/acs.jcim.5b00069}

\bibitem{1998cond.mat..3344W}
D.~{Wales}, J.~{Doye}, {Global Optimization by Basin-Hopping and the Lowest
  Energy Structures of Lennard-Jones Clusters Containing up to 110 Atoms},
  eprint arXiv:cond-mat/9803344\href {http://arxiv.org/abs/cond-mat/9803344}
  {\path{arXiv:cond-mat/9803344}}.

\bibitem{PhysRevLett.95.185505}
S.~T. Bromley, E.~Flikkema,
  \href{https://link.aps.org/doi/10.1103/PhysRevLett.95.185505}{Columnar-to-disk
  structural transition in nanoscale $({\mathrm{sio}}_{2}{)}_{N}$ clusters},
  Phys. Rev. Lett. 95 (2005) 185505.
\newblock \href {http://dx.doi.org/10.1103/PhysRevLett.95.185505}
  {\path{doi:10.1103/PhysRevLett.95.185505}}.
\newline\urlprefix\url{https://link.aps.org/doi/10.1103/PhysRevLett.95.185505}

\bibitem{A901227C}
S.~M.~Woodley, P.~D.~Battle, J.~D.~Gale, C.~Richard A.~Catlow,
  \href{http://dx.doi.org/10.1039/A901227C}{The prediction of inorganic crystal
  structures using a genetic algorithm and energy minimisation}, Phys. Chem.
  Chem. Phys. 1 (1999) 2535--2542.
\newblock \href {http://dx.doi.org/10.1039/A901227C}
  {\path{doi:10.1039/A901227C}}.
\newline\urlprefix\url{http://dx.doi.org/10.1039/A901227C}

\bibitem{JM9940400831}
T.~S. Bush, J.~D. Gale, C.~R.~A. Catlow, P.~D. Battle,
  \href{http://dx.doi.org/10.1039/JM9940400831}{Self-consistent interatomic
  potentials for the simulation of binary and ternary oxides}, J. Mater. Chem.
  4 (1994) 831--837.
\newblock \href {http://dx.doi.org/10.1039/JM9940400831}
  {\path{doi:10.1039/JM9940400831}}.
\newline\urlprefix\url{http://dx.doi.org/10.1039/JM9940400831}

\bibitem{1993JChPh..98.1372B}
A.~D. {Becke}, {A new mixing of Hartree-Fock and local density-functional
  theories}, jcp 98 (1993) 1372--1377.
\newblock \href {http://dx.doi.org/10.1063/1.464304}
  {\path{doi:10.1063/1.464304}}.

\bibitem{1996JChPh.105.9982P}
J.~P. {Perdew}, M.~{Ernzerhof}, K.~{Burke}, {Rationale for mixing exact
  exchange with density functional approximations}, jcp 105 (1996) 9982--9985.
\newblock \href {http://dx.doi.org/10.1063/1.472933}
  {\path{doi:10.1063/1.472933}}.

\bibitem{g09}
M.~J. Frisch, G.~W. Trucks, H.~B. Schlegel, G.~E. Scuseria, M.~A. Robb, J.~R.
  Cheeseman, G.~Scalmani, V.~Barone, B.~Mennucci, G.~A. Petersson,
  H.~Nakatsuji, M.~Caricato, X.~Li, H.~P. Hratchian, A.~F. Izmaylov, J.~Bloino,
  G.~Zheng, J.~L. Sonnenberg, M.~Hada, M.~Ehara, K.~Toyota, R.~Fukuda,
  J.~Hasegawa, M.~Ishida, T.~Nakajima, Y.~Honda, O.~Kitao, H.~Nakai, T.~Vreven,
  J.~A. Montgomery, {Jr.}, J.~E. Peralta, F.~Ogliaro, M.~Bearpark, J.~J. Heyd,
  E.~Brothers, K.~N. Kudin, V.~N. Staroverov, R.~Kobayashi, J.~Normand,
  K.~Raghavachari, A.~Rendell, J.~C. Burant, S.~S. Iyengar, J.~Tomasi,
  M.~Cossi, N.~Rega, J.~M. Millam, M.~Klene, J.~E. Knox, J.~B. Cross,
  V.~Bakken, C.~Adamo, J.~Jaramillo, R.~Gomperts, R.~E. Stratmann, O.~Yazyev,
  A.~J. Austin, R.~Cammi, C.~Pomelli, J.~W. Ochterski, R.~L. Martin,
  K.~Morokuma, V.~G. Zakrzewski, G.~A. Voth, P.~Salvador, J.~J. Dannenberg,
  S.~Dapprich, A.~D. Daniels, Ã.~Farkas, J.~B. Foresman, J.~V. Ortiz,
  J.~Cioslowski, D.~J. Fox, Gaussian∼09 {R}evision {E}.01, gaussian Inc.
  Wallingford CT 2009.

\bibitem{GU201529}
Y.~Gu, N.~Xu, M.~Lin, K.~Tan,
  \href{http://www.sciencedirect.com/science/article/pii/S2210271X15001425}{Structures,
  stabilities and properties of hollow (al2o3)n clusters (n=10, 12, 16, 18, 24
  and 33): Studied with density functional theory}, Computational and
  Theoretical Chemistry 1063 (2015) 29 -- 34.
\newblock \href
  {http://dx.doi.org/https://doi.org/10.1016/j.comptc.2015.03.027}
  {\path{doi:https://doi.org/10.1016/j.comptc.2015.03.027}}.
\newline\urlprefix\url{http://www.sciencedirect.com/science/article/pii/S2210271X15001425}

\bibitem{Pinto2004}
H.~P. Pinto, R.~M. Nieminen, S.~D. Elliott,
  \href{https://doi.org/10.1103/physrevb.70.125402}{Ab initiostudy
  of$\gamma$-al2o3surfaces}, Physical Review B 70~(12).
\newblock \href {http://dx.doi.org/10.1103/physrevb.70.125402}
  {\path{doi:10.1103/physrevb.70.125402}}.
\newline\urlprefix\url{https://doi.org/10.1103/physrevb.70.125402}

\bibitem{doi:10.1021/bk-1998-0677}
K.~K. Irikura, D.~J. Frurip,
  \href{https://pubs.acs.org/doi/abs/10.1021/bk-1998-0677}{Computational
  Thermochemistry}, American Chemical Society, Washington, DC, 1998.
\newblock \href
  {http://arxiv.org/abs/https://pubs.acs.org/doi/pdf/10.1021/bk-1998-0677}
  {\path{arXiv:https://pubs.acs.org/doi/pdf/10.1021/bk-1998-0677}}, \href
  {http://dx.doi.org/10.1021/bk-1998-0677} {\path{doi:10.1021/bk-1998-0677}}.
\newline\urlprefix\url{https://pubs.acs.org/doi/abs/10.1021/bk-1998-0677}

\bibitem{2010LNP...815..143M}
F.~J. {Molster}, L.~B.~F.~M. {Waters}, F.~{Kemper}, {The Mineralogy of
  Interstellar and Circumstellar Dust in Galaxies}, in: T.~{Henning} (Ed.),
  Lecture Notes in Physics, Berlin Springer Verlag, Vol. 815 of Lecture Notes
  in Physics, Berlin Springer Verlag, 2010, pp. 143--201.
\newblock \href {http://dx.doi.org/10.1007/978-3-642-13259-9_3}
  {\path{doi:10.1007/978-3-642-13259-9_3}}.

\end{thebibliography}

\section{Appendix}
\begin{table}[!h]
\caption{Thermodynamic quantities (temperature T, entropy S, heat capacity c$_P$, enthalpy 
change H(T)-H(0), enthalpy of formation dH$_f$ and Gibbs free energy of formation dG$_f$ of  8A. Units are K for T, JK$^{-1}$mol$^{-1}$ for S and c$_P$, and kJmol$^{-1}$ for H(T)-H(0),dH$_f$ and dG$_f$\label{td8a}.}
\begin{tabular}{l | r r r r r}
 T      &            S   &   C$_P$  &    H(T)-H(0) &   dH$_f$ (kJ/m)   &  dG$_f$       \\
 \hline 
0.00    &     0.000     &     0.000      &    0.000   &   -9114.469    &  -9114.469      \\
100.00  &   395.149     &   205.289      &    9.052   &   -9055.689    &  -8545.540      \\
200.00  &   619.018     &   459.222      &   42.819   &   -8992.442    &  -7887.085      \\
250.00  &   732.484     &   558.040      &   68.335   &   -8957.766    &  -7554.811      \\
298.15  &   837.577     &   634.649      &   97.118   &   -8922.095    &  -7234.368      \\
300.00  &   841.510     &   637.258      &   98.295   &   -8920.662    &  -7222.025      \\
350.00  &   944.617     &   699.536      &  131.778   &   -8881.499    &  -6889.260      \\
400.00  &  1041.339     &   748.222      &  168.022   &   -8840.751    &  -6556.870      \\
450.00  &  1131.760     &   786.395      &  206.426   &   -8798.843    &  -6225.041      \\
500.00  &  1216.237     &   816.552      &  246.528   &   -8756.149    &  -5893.967      \\
600.00  &  1369.237     &   860.003      &  330.512   &   -8669.469    &  -5234.332      \\
700.00  &  1504.114     &   888.813      &  418.046   &   -8582.487    &  -4578.544      \\
800.00  &  1624.174     &   908.693      &  507.979   &   -8496.474    &  -3927.026      \\
900.00  &  1732.067     &   922.903      &  599.597   &   -8412.296    &  -3280.074      \\
1000.00 &  1829.874     &   933.372      &  692.436   &   -8500.065    &  -2807.347      \\
1100.00 &  1919.221     &   941.289      &  786.186   &   -8415.723    &  -2164.472      \\
1200.00 &  2001.398     &   947.410      &  880.634   &   -8331.227    &  -1523.173      \\
1300.00 &  2077.429     &   952.234      &  975.625   &   -8246.644    &   -883.305      \\
1400.00 &  2148.144     &   956.101      & 1071.049   &   -8162.028    &   -244.770      \\
1500.00 &  2214.219     &   959.246      & 1166.822   &   -8077.399    &    392.473      \\
1600.00 &  2276.212     &   961.836      & 1262.880   &   -7992.845    &   1028.525      \\
1700.00 &  2334.590     &   963.995      & 1359.174   &   -7908.359    &   1663.434      \\
1800.00 &  2389.743     &   965.812      & 1455.667   &   -7823.970    &   2297.288      \\
1900.00 &  2442.004     &   967.356      & 1552.328   &   -7739.733    &   2930.067      \\
2000.00 &  2491.658     &   968.678      & 1649.131   &   -7655.602    &   3561.866      \\
2100.00 &  2538.948     &   969.819      & 1746.057   &   -7571.660    &   4192.683      \\
2200.00 &  2584.087     &   970.810      & 1843.090   &   -7487.851    &   4822.595      \\
2300.00 &  2627.261     &   971.677      & 1940.215   &   -7404.230    &   5451.636      \\
2400.00 &  2668.632     &   972.439      & 2037.422   &   -7320.775    &   6079.827      \\
2500.00 &  2708.343     &   973.113      & 2134.700   &   -7237.505    &   6707.158      \\
2600.00 &  2746.521     &   973.711      & 2232.042   &   -7154.387    &   7333.788      \\
2700.00 &  2783.279     &   974.245      & 2329.440   &   -7071.437    &   7959.704      \\ 
\end{tabular}
\end{table}

\begin{table}[!h]
\caption{Thermodynamic quantities (temperature T, entropy S, heat capacity c$_P$, enthalpy 
change H(T)-H(0), enthalpy of formation dH$_f$ and Gibbs free energy of formation dG$_f$ of isomer 8B. Units are K for T, JK$^{-1}$mol$^{-1}$ for S and c$_P$, and kJmol$^{-1}$ for H(T)-H(0),dH$_f$ and dG$_f$\label{td8b}.}
\begin{tabular}{l | r r r r r}
 T   &             &   C$_P$  &    H(T)-H(0) &   dH$_f$    &  dG$_f$       \\
 \hline
   0.00 &     0.000     &     0.000   &       0.000   &   -9100.292    &  -9100.292    \\
 100.00 &   387.576     &   197.744   &       8.595   &   -9041.969    &  -8531.063    \\
 200.00 &   607.633     &   456.565   &      41.841   &   -8979.243    &  -7871.609    \\
 250.00 &   720.715     &   557.171   &      67.273   &   -8944.651    &  -7538.753    \\
 298.15 &   825.748     &   634.774   &      96.041   &   -8908.995    &  -7217.741    \\ 
 300.00 &   829.682     &   637.411   &      97.217   &   -8907.563    &  -7205.378    \\
 350.00 &   932.857     &   700.208   &     130.723   &   -8868.377    &  -6872.022    \\
 400.00 &  1029.685     &   749.124   &     167.007   &   -8827.589    &  -6539.047    \\
 450.00 &  1120.218     &   787.372   &     205.458   &   -8785.634    &  -6206.638    \\
 500.00 &  1204.798     &   817.523   &     245.610   &   -8742.890    &  -5874.989    \\
 600.00 &  1357.967     &   860.870   &     329.685   &   -8656.119    &  -5214.220    \\
 700.00 &  1492.968     &   889.549   &     417.300   &   -8569.056    &  -4557.311    \\
 800.00 &  1613.118     &   909.310   &     507.301   &   -8482.975    &  -3904.682    \\
 900.00 &  1721.078     &   923.422   &     598.975   &   -8398.741    &  -3256.628    \\
1000.00 &   1818.935    &    933.812  &      691.862  &    -8486.462   &   -2782.805   \\
1100.00 &   1908.322    &    941.664  &      785.653  &    -8402.079   &   -2138.839   \\
1200.00 &   1990.528    &    947.733  &      880.135  &    -8317.549   &   -1496.451   \\
1300.00 &   2066.583    &    952.515  &      975.157  &    -8232.935   &    -855.496   \\
1400.00 &   2137.318    &    956.347  &     1070.607  &    -8148.293   &    -215.879   \\
1500.00 &   2203.409    &    959.462  &     1166.402  &    -8063.642   &     422.445   \\
1600.00 &   2265.415    &    962.028  &     1262.481  &    -7979.067   &    1059.578   \\
1700.00 &   2323.804    &    964.166  &     1358.794  &    -7894.562   &    1695.567   \\
1800.00 &   2378.967    &    965.966  &     1455.303  &    -7810.157   &    2330.498   \\
1900.00 &   2431.236    &    967.495  &     1551.978  &    -7725.906   &    2964.353   \\
2000.00 &   2480.896    &    968.804  &     1648.795  &    -7641.761   &    3597.231   \\
2100.00 &   2528.192    &    969.934  &     1745.733  &    -7557.807   &    4229.124   \\
2200.00 &   2573.337    &    970.915  &     1842.776  &    -7473.988   &    4860.108   \\
2300.00 &   2616.515    &    971.774  &     1939.912  &    -7390.356   &    5490.226   \\
2400.00 &   2657.889    &    972.528  &     2037.128  &    -7306.892   &    6119.493   \\
2500.00 &   2697.604    &    973.195  &     2134.415  &    -7223.613   &    6747.897   \\
2600.00 &   2735.785    &    973.787  &     2231.764  &    -7140.488   &    7375.601   \\
2700.00 &   2772.546    &    974.315  &     2329.170  &    -7057.530   &    8002.590   \\
\end{tabular}
\end{table}

\begin{longtable}[!h]{r r r r r r}
\caption{Vibration frequencies (in cm$^{-1}$) of structures 8A, 8B, and 8G, respectively, as 
calculated on the B3LYP/6-311+G(d) level of theory.\label{frequen}}
%\begin{tabular}{r r r r r r}
\\
8A \\
\hline
   90.2094 &  103.6610 &  113.1299 & 113.7897 &  126.2836  & 132.1154  \\
  149.5395 &  156.1298 &  158.4286 & 161.9488 &  174.7097  & 178.0820  \\
  184.1340 &  190.3925 &  204.5866 & 208.6846 &  221.3737  & 224.7862  \\
  234.3400 &  239.6439 &  242.4419 & 247.5123 &  258.1091  & 263.3042  \\
  270.1901 &  274.4194 &  278.7248 & 281.9794 &  282.3983  & 287.7413  \\
  288.8951 &  296.8160 &  298.2015 & 304.0709 &  313.6256  & 317.7489  \\
  319.7042 &  323.7852 &  336.3023 & 341.3491 &  348.0401  & 349.9338  \\
  358.6037 &  368.5844 &  372.0300 & 373.5352 &  377.7690  & 385.1710  \\
  391.3964 &  396.4796 &  397.9887 & 408.3760 &  411.1994  & 415.4799  \\
  429.5667 &  445.8082 &  454.5025 & 457.4732 &  468.7610  & 470.0209  \\
  500.3456 &  528.1360 &  529.2779 & 531.0116 &  543.8943  & 546.9150  \\
  548.7668 &  567.2123 &  569.4951 & 579.9749 &  588.3215  & 596.2581  \\
  604.8862 &  610.0111 &  619.6286 & 621.5216 &  628.7537  & 642.1270  \\
  644.2739 &  646.3816 &  657.8524 & 673.2383 &  677.6540  & 687.9148  \\
  695.1378 &  700.3167 &  712.7577 & 718.9825 &  728.6398  & 744.8394  \\
  750.7953 &  776.8708 &  778.8075 & 785.2246 &  789.7197  & 791.7034  \\
  814.7539 &  819.7563 &  830.9581 & 833.2751 &  841.4871  & 849.4601  \\
  858.8782 &  873.2397 &  876.3019 & 884.4792 &  902.1171  & 903.8100  \\
  910.3530 &  912.0732 &  927.6888 & 932.0930 &  977.5391  & 998.7102  \\
\hline
8B \\
\hline
 91.5358 & 102.2772 & 127.9649 & 131.8299 & 138.9168 & 148.8552 \\
162.9458 & 164.5536 & 170.5986 & 182.0207 & 182.4413 & 191.8727 \\
195.4770 & 201.6917 & 209.1333 & 214.7327 & 219.3009 & 226.7387 \\ 
237.6343 & 244.6223 & 257.7512 & 260.7836 & 263.6712 & 269.5421 \\  
272.3050 & 275.5059 & 276.6740 & 281.4691 & 284.9971 & 290.0597 \\
300.2656 & 305.9572 & 310.1684 & 314.8178 & 321.0454 & 323.5269 \\
328.9138 & 334.0525 & 335.7036 & 341.8931 & 343.0069 & 353.5171 \\
363.3137 & 371.8111 & 374.0156 & 380.3567 & 381.6630 & 386.0263 \\  
396.6048 & 404.7217 & 407.9325 & 416.3372 & 421.6654 & 427.1868 \\
438.9616 & 447.2230 & 454.3461 & 465.7418 & 472.3817 & 481.2743 \\
495.0933 & 504.2488 & 517.5900 & 521.9706 & 540.6087 & 550.9677 \\ 
557.2089 & 561.7813 & 570.0408 & 573.0887 & 580.7838 & 596.3860 \\
600.3627 & 611.0725 & 613.8789 & 617.0713 & 623.8788 & 627.7754 \\
633.1866 & 644.1204 & 656.4854 & 665.7699 & 674.8369 & 683.6253 \\ 
689.7383 & 701.3332 & 706.8205 & 712.2638 & 724.6251 & 735.9241 \\  
744.9986 & 750.8257 & 763.5632 & 768.4292 & 773.4613 & 788.8824 \\
799.3710 & 809.5934 & 813.1208 & 822.8043 & 824.5475 & 837.2392 \\ 
841.2173 & 851.2534 & 871.5892 & 873.8353 & 883.6798 & 893.3444 \\  
908.3880 & 922.3289 & 938.0098 & 942.0166 & 966.3894 & 987.5792 \\
\hline
8G \\
\hline
122.3280 & 122.3759 & 129.8718 & 131.3459 & 135.3512 & 135.3567 \\
160.7597 & 160.7656 & 183.8549 & 186.7616 & 187.1831 & 204.2287 \\ 
204.2723 & 217.8290 & 220.2045 & 229.5106 & 229.5346 & 231.3625 \\   
231.8471 & 239.0043 & 239.1124 & 252.2576 & 269.6685 & 272.3767 \\
281.5665 & 282.3724 & 282.3769 & 282.9120 & 286.2491 & 291.0896 \\
291.1085 & 302.0007 & 318.3927 & 321.6422 & 321.6754 & 335.8252 \\
335.8279 & 340.0566 & 346.0378 & 356.3743 & 356.3883 & 357.4173 \\  
382.5699 & 387.4643 & 389.4900 & 395.7846 & 395.7900 & 428.6817 \\
429.0807 & 437.5835 & 437.6173 & 465.4728 & 468.8242 & 468.8270 \\
479.5008 & 479.5210 & 496.0409 & 499.3839 & 511.9482 & 520.9266 \\ 
527.3433 & 527.8370 & 532.0045 & 532.0267 & 552.8827 & 557.4152 \\   
558.3805 & 559.2673 & 563.9438 & 563.9612 & 569.7835 & 569.8415 \\
590.0452 & 590.1963 & 590.2086 & 615.5018 & 616.9330 & 616.9430 \\
631.5871 & 634.1916 & 640.4553 & 640.4813 & 643.1729 & 649.5379 \\
662.7197 & 662.7357 & 664.9720 & 669.2429 & 701.3175 & 701.3529 \\   
713.2892 & 726.5958 & 728.8081 & 737.7250 & 737.7345 & 757.0507 \\
768.7273 & 768.7298 & 783.9297 & 788.5174 & 816.6344 & 822.2092 \\ 
822.2348 & 850.9110 & 853.7232 & 853.8241 & 857.5930 & 865.4159 \\ 
874.7006 & 874.7399 & 898.7439 & 928.8732 & 928.9094 & 952.6200 \\
%\end{tabular}
\end{longtable}

\end{document}